\documentclass[a4paper,10pt]{article}
\usepackage[a4paper, margin=1in]{geometry}
\usepackage{graphicx}
\usepackage{multirow}
\usepackage{subcaption} 
\usepackage{booktabs}
\usepackage{mathtools}
\usepackage{amsthm}
\usepackage{amsfonts}
\usepackage{tabularx} 
\usepackage{alphabeta}

\usepackage[hidelinks]{hyperref}
\usepackage{xcolor}

\newcount\Comments  
\definecolor{darkgreen}{rgb}{0,0.5,0}
\newcommand{\kibitz}[2]{\ifnum\Comments=0\textcolor{#1}{#2}\fi}
\newcommand{\Yanbing}[1]{\kibitz{blue}      {[YW: #1]}}

\newcommand{\q}{\mathbf{q}}
\newcommand{\thetacf}{\theta_{\textrm{CF}}}
\newcommand{\thetalc}{\theta_{\textrm{LC}}}

\title{Calibrating microscopic traffic models with macroscopic data}

\date{}

\author{Yanbing Wang
\thanks{Arizona State University, Tempe AZ, USA; Email: yanbing.wang@asu.edu} 
\and Felipe de Souza
\thanks{Argonne National Laboratory, Lemont IL, USA}
\and Dominik Karbowski\footnotemark[2]}

\begin{document}
\maketitle

\section{Abstract}
Traffic microsimulation is a crucial tool that uses microscopic traffic models, such as car-following and lane-change models, to simulate the trajectories of individual agents. This digital platform allows for the assessment of the impact of emerging technologies on transportation system performance. While these microscopic models are based on mathematical structures, their parameters must be fitted to real-world data through a process called model calibration. Despite extensive studies on calibration, the focus has predominantly been on fitting microscopic data, such as trajectories, rather than evaluating how well the models reproduce macroscopic traffic patterns, such as congestion, bottlenecks, and traffic waves. In this work, we address this gap by calibrating microscopic traffic flow models using macroscopic (aggregated) data, which is more readily accessible. We designed a SUMO-in-the-loop calibration framework with the goal of replicating observed macroscopic traffic features. To assess calibration accuracy, we developed a set of performance measures that evaluate the models' ability to replicate traffic states across the entire spatiotemporal domain and other qualitative characteristics of traffic flow. The calibration method was applied to both a synthetic scenario and a real-world scenario on a segment of Interstate 24, to demonstrate its effectiveness in reproducing observed traffic patterns.

\hfill\break%
\noindent\textit{Keywords}: Traffic Microsimulation, Model Calibration, Traffic Flow
\newpage

\section{Introduction}
\subsection{Motivation}
The motivation for studying microscopic model calibration to reproduce macroscopic observations arises from the need to overcome the limitations of traditional calibration approaches. While detailed trajectory data (e.g., NGSIM~\cite{NGSIM2006}, HighD~\cite{krajewski2018highd}, openACC~\cite{makridis2020openacc}) is beneficial for calibrating microscopic models, such as car-following and lane-change models, the collective driving behaviors and traffic patterns generated from these calibrated parameters may not always be accurate. Montanino and Punzo~\cite{montanino2015trajectory} showed that even with individually calibrated car-following and lane-changing parameters, simulations failed to reproduce key features like congestion propagation. Similarly, studies on macroscopic traffic model calibration~\cite{MOHAMMADIAN2021132} found that a well-calibrated model could still fail to reproduce observed travel times, congestion patterns, and traffic flow features. These findings raise a fundamental question for traffic model calibration: \textit{what constitutes good model calibration, and how can we design a calibration framework that incorporates key observations of traffic characteristics?}

To address this question, this paper presents a SUMO-in-the-loop calibration procedure utilizing macroscopic traffic measurements, typically obtained from detector loops or similar roadside sensors. SUMO~\cite{SUMO2018} is an open-source traffic micro-simulation software.
Through a synthetic scenario and a real-world freeway scenario, we explore (1) which measurements contribute most to calibration performance, (2) which model features (car-following and/or lane-change) best capture observed traffic patterns, and (3) how calibrated models perform with respect to other performance measures, such as the ability to reproduce macroscopic traffic patterns. The calibrated SUMO model can generate detailed simulated trajectories for every vehicle and their interactions while ensuring that overall macroscopic features resemble real-world observations. This calibrated microsimulation platform is crucial for testing and quantifying the performance of emerging vehicle control and traffic management strategies.

\subsection{Contributions}
We highlight the following contributions in this paper:
\begin{enumerate}
    \item We designed a SUMO-in-the-loop framework to calibrate traffic microsimulation models using only aggregated (macroscopic) traffic measurements, without relying on individual trajectories. This framework is demonstrated through both a synthetic scenario and a real-world scenario with field data to reproduce observed traffic patterns.
    \item We provided an analysis of traffic measurements (flow, occupancy, and speed) and model features (car-following and/or lane-changing) on the calibration performance. Additionally, we discuss other performance measures of the calibrated models, such as the ability to reproduce congestion patterns.
    \item We released a repository accompanying the calibration procedure and analysis in this paper to support reproducible research. [Link removed for review process]
\end{enumerate}

The rest of the paper is organized as following: Section~\ref{sec:lit_review} provides an overview of traffic flow model calibration studies, their challenges, and new perspectives; Section~\ref{sec:method} presents the microsimulation models and parameters considered in this work, as well as the calibration framework. Section~\ref{sec:experiments} discusses the calibration performance in both a synthetic corridor and a real-world freeway corridor. Finally we highlight the results and future directions in Section~\ref{sec:conclusion}.

\section{Literature Review}
\label{sec:lit_review}
The field of traffic modeling has made significant strides with the development of traffic sensing devices and the increased availability of measurement data. Greenshield's pioneering work in the 1930s~\cite{greenshields1935study} laid the groundwork with a simple model relating vehicle flow and travel speeds, establishing a scientific approach to traffic analysis. Macroscopic traffic flow models have since then evolved, which draws parallels with hydrodynamics. Key contributions include Lighthill and Whitham's kinematic wave theory~\cite{lighthill1955kinematic} and Richards' shockwave theory~\cite{richards1956shock}. These theories were further developed by Newell's simplified kinematic wave theory~\cite{newell1993simplified1,newell1993simplified2,newell1993simplified3}, Prigogine and Herman's gas-kinetic theories~\cite{herman1979two}, and higher-order flow models~\cite{payne1971models,daganzo1995requiem,aw2000resurrection}. 

Microscopic traffic models, focusing on individual driver behavior, are crucial for linking driving behavior to overall traffic characteristics. Notable models include the Stimulus Response Model~\cite{nagel1992cellular}, Collision Avoidance Models~\cite{newell1961nonlinear,gipps1981behavioural,kometani1958stability}, the Optimal Velocity Model~\cite{bando1995phenomenological}, the Intelligent Driver Model~\cite{treiber2000congested}, and Cellular Automata Models~\cite{nagel1992cellular}. These models describe car-following behaviors and have applications in human driver analysis as well as adaptive cruise control (ACC) systems. Additionally, models focusing on lane-changing and gap acceptance behaviors~\cite{treiber2009modeling,laval2006lane} have been instrumental in traffic microsimulation, control design, and stability analysis~\cite{wang2014rolling,swaroop1996string,li2017distributed,oncu2014cooperative,liang1999optimal,naus2010string,gunter2020commercially}.

The calibration of such models is essential to ensure these models faithfully represent real-world conditions. Calibration adjusts model parameters to match observed traffic patterns, which is now supported by the availability of detailed traffic measurements and datasets. One of the earliest form of traffic measurements are from loop detectors, which measures aggregated data such as flow and occupancy within a certain time intervals. These macroscopic measurements have supported calibration of macroscopic traffic flow models~\cite{ngoduy2012calibration,mudigonda2015robust,spiliopoulou2017macroscopic,brockfeld2005calibration,MOHAMMADIAN2021132} and model-based traffic state estimation~\cite{WANG2005141,VANLINT200814078,hegyi2006,BLANDIN20121421,Doucet11atutorial,chen2003,mihaylova2007,Mihaylova2012,wright2016,wang2017trafficestimators,Polson2018,SEO2017128}.

Calibrating microscopic models presents significant challenges due to the need for detailed datasets, including individual trajectories with high-frequency information on acceleration, speed, and lane changes. Historically, the lack of such comprehensive data has posed difficulties, particularly as microscopic models require more input parameters compared to macroscopic models. In the early days, calibration methods for microscopic models rarely had access to detailed vehicle trajectories. Instead, these methods relied on aggregate data from loop detectors and assumptions based on traffic engineering experience. For instance, the guidelines proposed by Dowling et al.~\cite{dowling2004guidelines} acknowledged the advantages of floating car data but also recognized its unavailability. They proposed a flexible three-step calibration method (capacity, route choice, and system performance), with capacity calibration ideally based on field measurements from point sensors like loop detectors, which provide only aggregated data. When such data were unavailable, they suggested estimating capacity from the Highway Capacity Manual. For congested conditions, they claim “it may require 10 or more runs”.

Similar procedures from that time, such as those by~\cite{chu2003calibration, liu2006streamlined}, also relied on aggregate data and limited individual vehicle data. Manual calibration was common due to limited toolsets and the long runtime of microscopic simulations. However, some automatic calibration approaches using simulation-based algorithms demonstrated potential, as highlighted in~\cite{hollander2008principles}. These automatic methods reduced the number of variables to be calibrated by leveraging prior calibration efforts and focusing on local parameters first, ensuring fewer variables in the final stage.


As the field progressed, the availability of microscopic traffic data (e.g., detailed trajectories) along with aggregated data increased substantially, which supported trajectory data-based microsimulation model calibration. According to a recent review~\cite{LI2020225}, the problem of microsimulatino model calibration can be categorized into four types~\cite{panwai2005comparative,Punzo2012CanRO,punzo2015do,LI2017headway}. Briefly put, \textit{type I} calibration views the calibration problem as a likelihood estimation problem where the distribution of the parameter likelihood is calculated for the future time step based on historical driving data (e.g., \cite{panwai2005comparative, Treiber2013, hoogendoorn2010generic,TREIBER2013922,punzo2015do,PunzoModelCalibration2005, wang2020estimating, zhang2024bayesian}). \textit{Type II} calibration directly uses a global search to find the best-fit parameters for which the simulated complete trajectory most closely represent the observed trajectory (e.g., \cite{ma2002genetic,wang2010usingTraj, ciuffo2014nofree,Papathanasopoulou2015towards,gunter2019modelcomparison, wang2022identifiability}). \textit{Type III} calibration considers the long-range interactions amongst vehicles within a platoon (e.g., \cite{LAVAL2014228, Kurtc2015calibratingLocal,HE2015185}), and \textit{type IV} calibration relies on mesoscopic or macroscopic traffic flow patterns such as the headway distributions (e.g., \cite{JIN2009318,LI2017headway}). Other types of microsimulation calibration focus on innovative metrics, such as the capacity drop phenomena~\cite{de2022unveiling} and energy and mobility metrics~\cite{long2024bi}.

Detailed trajectory data is undoubtedly advantageous for calibrating microscopic models, which require a significant number of parameters to represent the collective driving behavior. However, Montanino and Punzo~\cite{montanino2015trajectory} demonstrated that relying solely on microscopic calibration may not suffice. They calibrated car-following and lane-changing parameters from NGSIM data for each vehicle, but a "trace-driven" simulation with these parameters resulted in macroscopic quantities that differed significantly from observed data. Key features, such as congestion propagation, were not accurately reproduced. The reasons included: (1) trace-driven simulation accumulates errors, and upstream demand was not represented in the data; (2) simulated on-ramp merging behavior did not realistically reflect the extent seen in the NGSIM dataset. This work underscores the challenges in calibrating microsimulation models and raises the fundamental question: "Is calibrating against disaggregated data the correct way to reproduce emergent macroscopic patterns?" In a later study~\cite{punzo2020two}, the same authors proposed a two-level probabilistic approach to account for various sources of heterogeneity in calibration.

The work by Montanino and Punzo~\cite{montanino2015trajectory} highlights a fundamental challenge in calibration: the insufficiency of relying solely on the goodness of fit to individual trajectories, and the necessity of considering other performance measures essential for assessing a model's predictive capabilities. This challenge also applies to calibrating continuum models (macroscopic traffic flow models)~\cite{MOHAMMADIAN2021132}. This study shows that effective calibration must consider not only the goodness of fit—the difference between model-simulated measurements and real-world observations, but also a set of operational measures such as delay, travel time, and congestion patterns, as well as the model's ability to reproduce complex traffic phenomena like oscillations and capacity drops. Some models may achieve a good fit with small calibration errors but still exhibit large operational errors. Therefore, a well-rounded assessment criterion is necessary for models to be suited for real-world applications.

\section{Methodology}
\label{sec:method}
This work focuses on calibrating driving behavior parameters only, including car-following parameters $\thetacf$ and lane-change parameters $\thetalc$, and they are summarized in Table~\ref{tab:parameters}. 

\subsection{Preliminary: Intelligent Drivers Model}
The car-following model used in this study is the well-known and one of the most commonly used Intelligent Driver Model (IDM)~\cite{treiber2000congested} that represents the car-following behavior of a realistic human driver, such as asymmetric accelerations and decelerations. The acceleration of the follower $\dot{v}(t)$ is expressed as an ordinary differential equation (ODE) with respect to the follower's speed $v(t)$, the relative distance $s(t)$ and relative speed $\Delta v(t)$ to the leader:
\begin{equation}
\label{eq:Enhanced_IDM}
\dot{v}(t) = a\left[1-\left(\dfrac{v(t)}{v_f}\right)^{\delta}-\left(\dfrac{s^*(v(t),\Delta v(t))}{s(t)}\right)^2\right]
\end{equation}
where the desired space gap $s^*$ is defined as:
\begin{equation}
    s^*(v(t),\Delta v(t)) = s_j + v(t) \tau + \dfrac{v(t)\Delta v}{2\sqrt{ab}}.
\end{equation}
where $v_f$ is the freeflow speed,  $s_j$ the jam space-gap to the leader, $\tau$ the desired time headway, $a$ maximum acceleration and $b$ the desired deceleration. The exponent parameter $\delta$ is usually set as 4~\cite{treiber2000congested}. The five free car-following parameters $\theta_{\textrm{CF}} = [s_j, v_f, \tau, a, b]$ are subject to calibration. The IDM is a collision-free car-following model that can capture complex traffic phenomena such as the formation and dissipation of traffic waves. While IDM allows deceleration rates higher than the comfortable deceleration~\cite{treiber2013car}, the maximum deceleration is enforced in SUMO as the emergency deceleration ($8m/s^2$).

\subsection{Preliminary: Lane-Change Strategies}

SUMO features a lane-changing model in which driver-vehicle units perform different lane-changing actions based on their motivations \cite{erdmann2015sumo}. The four possible motivations are strategic, cooperative, tactical, and regulatory lane changes. Like other lane-changing models, SUMO's model considers the feasibility of the maneuver and a gap acceptance rule that ensures drivers avoid being too close to vehicles in the target lane. Gap acceptance bounds are also influenced by car-following parameters.

The simplest case is tactical lane changes, where drivers change lanes to gain a speed advantage by avoiding a slow leader in the current lane. The willingness to perform such changes is controlled by the lcKeepRight and lcSpeedGain parameters (see Table \ref{tab:parameters}).

Complex maneuvers occur around merge and diverge areas, where vehicles must change lanes to stay on their intended route. This is known as strategic lane changing. The model calculates the number of required lane changes based on the current link and lane and the intended route. The likelihood of performing these maneuvers increases with the number of required lane changes, the distance, and the approaching speed of the next required lane change. When no acceptable gap is available in the target lane, the ego vehicle and the leader and follower in the target lane undertake actions to complete the maneuver. The urgency and aggressiveness of this maneuver are controlled by the lcStrategic and lcAssertive parameters.

When vehicles struggle to find gaps in the adjacent lane for strategic lane changes, vehicles in the adjacent lane may also change lanes if possible to facilitate the maneuver. This behavior is controlled by the lcCooperative parameter.

The lane-changing parameters to be calibrated in this work are $\theta_{\textrm{LC}}$ = [\textrm{lcStrategic, lcCooperative, lcAssertive, lcSpeedGain, lcKeepRight}]. For detailed information on the parameters and the lane-changing model, please refer to the SUMO user manual and the lane-changing model documentation~\cite{erdmann2015sumo}.

\begin{table}[!ht]
\resizebox{\textwidth}{!}{%
\begin{tabular}{@{}lcccc@{}} 
\toprule
Model features                 & Parameter name & Description & Ground truth & Default \\ \midrule
\multirow{5}{*}{Car-following} & $v_f$         &    Maximum speed [m/s]   & 30.55 & 32.0 \\
                               & $s_j$          &  minimum space gap [m]     & 2.5 & 2.5 \\
                               & $\tau $             &   desired time headway [sec]     & 1.4 & 1.0  \\
                               & $a$              &    maximum acceleration [m/s$^2$]     & 1.5 &  2.6 \\
                               & $b$              &   maximum deceleration [m/s$^2$]      & 2.0 &  4.5  \\ \midrule
\multirow{6}{*}{Lane-change}   & lcStrategic    &  The eagerness for performing strategic lane changing     & 1.0 &   1.0     \\
                               & lcCooperative           &     The willingness for performing cooperative lane changing.   &1.0  &   1.0    \\
                               &  lcAssertive        &  Willingness to accept lower front and rear gaps on the target lane.     & 0.5 &    1.0     \\
                               &  lcSpeedGain    &   The eagerness for performing lane changing to gain speed.         & 1.0 &  1.0  \\
                               &   lcKeepRight       &    The eagerness for following the obligation to keep right.    & 0.5 &   1.0   \\ 
                               \bottomrule
\end{tabular}}
\caption{Parameters subject to calibration, their descriptions, and values}
\label{tab:parameters}
\end{table}

\subsection{Optimization Formulation}
 The calibration process is an optimization of finding the best-fit $\thetacf$ and $\thetalc$ with SUMO-in-the-loop. The goal is to find those parameters such that the simulated measurements from SUMO loop detectors best match the data (either synthetic or from field sensors).

The general optimization setup is the following:
\begin{equation}
\label{eq:opt}
\begin{aligned}
\theta^* &= \underset{\theta \in \Theta}{\text{argmin}}\ f\left(Z^{\text{obs}, \omega}, Z^{\text{sim}, \omega}(\theta, \Gamma)\right),
\end{aligned}
\end{equation}
where $Z^{\text{obs}, \omega}, Z^{\text{sim}, \omega}$ are vectors of the observed (aggregated) traffic measurements from stationary detectors and the corresponding (aggregated) simulated measurements at the same spatial and temporal domain, defined by $\omega = [0,l] \times [0,t]$.
The objective function $f$ measures the discrepancy between the observation and the simulated measurements (or a measure of goodness of fit). It is evaluated with a micro-simulation inside, which is not available in closed-form.
The decision variable $\theta \in \{\theta_{\textrm{CF}}, \theta_{\textrm{LC}}\} $ is a subset of all the driving behavioral parameters to be calibrated, bounded by $\Theta$; $\Gamma$ includes all the endogenous and exogenous simulation configurations such as network topology, routes, upstream and ramp demands, which are prescribed for each scenario. Note that due to nonlinearlity of the micro-simulation model and the non-differentiability of the objective function with respect to the parameters, the optimization problem can have multiple local minima. Therefore in this paper, a global optimization technique, differential evolution algorithm (DE)~\cite{storn1997differential} is applied to increase the chance of exploring solutions outside of the local minima. For a discussion on optimization algorithm for microsimulation model calibration, please refer to~\cite{hollander2008principles}.

In this work, the fixed-location measurements $Z^{\text{obs}, \omega}$ include aggregated measure of flow, speed and occupancy at specific locations of each lane. We choose a  goodness of fit in the objective function as the element-wise root mean squared error (RMSE) between observed and simulated measurements:
\begin{equation}
\label{eq:opt_rmse}
f\left(Z^{\text{obs}, \omega}, Z^{\text{sim}, \omega}\right) \coloneqq \left( \dfrac{1}{|\omega|}\sum_{i=0}^t \sum_{j=0}^l \left(\zeta_{i,j}^{\text{obs}, \omega} - \zeta_{i,j}^{\text{sim}, \omega} \right)^2 \right)^{1/2},
\end{equation}

where $ |\omega| = t\times l$ is the number of measurement points.
Later in Section~\ref{sec:experiments}, we discuss the impact of specific measurements on the calibration performance. 

\section{Experiments}
\label{sec:experiments}
In this section, we perform model calibration on two scenarios. The first is a synthetic freeway on-ramp merging scenario, with measurements for parameter calibration simulated using SUMO. The second one replicates a real-world scenario from Interstate 24 post-miles 54.1 to 57.6. This segment includes two on-ramps, one off-ramp, and various merging zones. Real-world radio detection systems (RDS) data obtained from the Tennessee Department of Transportation is used for model calibration. The calibrated models are tested using a validation dataset consisting of higher-resolution data with macroscopic traffic quantities spanning the entire spatial-temporal domain. The simulated traffic patterns are compared with the validation dataset to evaluate performance. A summary of the two experiments is shown in Table~\ref{tab:two_experiments}.

\begin{table}[]
\resizebox{\textwidth}{!}{%
\begin{tabular}{@{}lllllll@{}}
\toprule
Corridor         & Spatial range & Temporal range & Measurements & Validation data        & Ramps                     & Lanes     \\ \midrule
Synthetic & 1300 m       & 8 min         & Simulated loop detectors    & Simulated trajectories     & 1 on-ramp                 & 2-3 lanes \\
I-24      & 3.5 miles    & 3 hours       & RDS data     & RDS with ASM*           & 2 on-ramps, 1 off-ramp & 4-6 lanes \\ \bottomrule
\end{tabular}}
\caption{Summary of experiments setup. *ASM: adaptive smoothing method.}
\label{tab:two_experiments}
\end{table}

\subsection{Numerical Experiment: On-Ramp Merging}

The purpose of this numerical experiment is to provide a deterministic and ground truth micro-simulation scenario, with known properties, to test the ability of the proposed method on reproducing the observed traffic characteristics. The simulation setup consists of the network geometry, demands (flows that enters the network), and fleet properties such as their car-following and lane-change parameters. 

\begin{figure}[!ht]
  \centering
  \includegraphics[width=\textwidth]{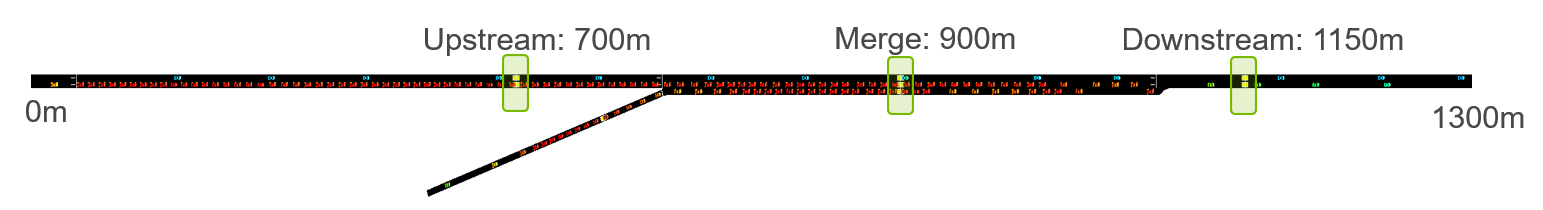}
  \caption{Synthetic corridor setup in SUMO. Three loop-detectors are indicated by the green boxes.}
  \label{fig:synth_corridor}
\end{figure}

\subsubsection{Experiment Setup}
We consider a short freeway segment of 1300m long and with an on-ramp merging zone (Figure~\ref{fig:synth_corridor}). The mainline has 2 lanes, while the merging zone (from 800-1100m) has 3 lanes. A bottleneck at the merging zone is activated when the combined flow from the upstream and the on-ramp exceeds its capacity.
 The ground truth and the default (uncalibrated) driving behavior parameters are listed in Table~\ref{tab:parameters}. Note that the default scenario differs from the ground truth scenario only by these parameters, which are subject to calibration. All else (network geometry and the demands) are kept equal.
 
 Three loop detectors are located along the segment at position 700m (upstream of merge), 900m (merging zone) and 1150m (downstream of merge), respectively. These loop detector measurements are obtained from SUMO (e1Detector element) that include lane-specific, aggregate measurements of flow $\q$ (in vehicles per hour or vph), speed $\mathbf{v}$ (in miles per hour or mph) and occupancy $\mathbf{o}$ (in \%). All measurements are aggregated in 50-sec windows.

For each target measurement, we consider three model features: (1) calibrating only $\thetacf$, (2) calibrating only $\thetalc$, and (3) calibrating both. This study results in nine different combinations of experiments, and they are labeled in Table~\ref{tab:exp_labels}. Exp. "x.a" uses flow data only, "x.b" uses velocity data only, and "x.c" uses occupancy data only, and The purpose is to test the impact of model features (car-following and lane-change), as well as measurements, on the ability to reproduce macroscopic traffic features. 

For example, one instance of the optimization problem~\eqref{eq:opt_rmse} corresponds to Exp.1.a. - calibrating car-following model parameters only against flow data with lane-change parameters fixed at default, is written as:
\begin{equation}
\label{eq:opt_1a}
\begin{aligned}
\thetacf^* &= \underset{\thetacf \in \Theta}{\text{argmin}} \left( \dfrac{1}{|\omega|}\sum_{i=0}^t \sum_{j=0}^l \left(q_{i,j}^{\text{obs}, \omega} - q_{i,j}^{\text{sim}, \omega}(\thetacf,\thetalc^*) \right)^2 \right)^{1/2},
\end{aligned}
\end{equation}
and similarly, Exp 2.b. can be explicitly written as:
\begin{equation}
\label{eq:opt_2b}
\begin{aligned}
\thetalc^* &= \underset{\thetalc \in \Theta}{\text{argmin}} \left( \dfrac{1}{|\omega|}\sum_{i=0}^t \sum_{j=0}^l \left(v_{i,j}^{\text{obs}, \omega} - v_{i,j}^{\text{sim}, \omega}(\thetacf^*,\thetalc) \right)^2 \right)^{1/2},
\end{aligned}
\end{equation}
Ditto for others. 

\begin{table}[]
\begin{tabular}{@{}lllll@{}}
\toprule
                                                 &                   & \multicolumn{3}{l}{Measurements used for calibration}   \\ \cmidrule(l){3-5} 
Parameters to be calibrated                      & No. of parameters & Flow $\q$ & Speed $\mathbf{v}$ & Occupancy $\mathbf{o}$ \\ \midrule
$\theta_{\textrm{CF}}$                           & 5                 & 1.a       & 1.b                & 1.c                    \\
$\theta_{\textrm{LC}}$                           & 5                 & 2.a       & 2.b                & 2.c                    \\
$\theta_{\textrm{CF}}$ and $\theta_{\textrm{LC}}$ & 10                & 3.a       & 3.b                & 3.c                    \\ \bottomrule
\end{tabular}
\caption{Experiment labels based on the combination of decision variables and measurements}
\label{tab:exp_labels}
\end{table}

\subsubsection{Measure of Performance}
As mentioned in~\cite{MOHAMMADIAN2021132}, various benchmarking metrics should be considered to assess the calibration performance. The calibration error directly corresponds to the objective function value, i.e., the discrepancy between the measurements and the simulated measurements. However the calibrated model should also produce sensible traffic characteristics, and therefore the ability to recover the entire spatiotemporal traffic patterns need also to be considered.

Therefore we evaluate the results using two sets of errors: RMSE at detector locations, and RMSE of macroscopic fields. The RMSE at detector locations is calculated from the difference between the measurements and the simulated measurements at the selected detector locations (see Figure~\ref{fig:synth_corridor}), and it directly corresponds to the objective function values when the optimization terminates: \Yanbing{change to RMSPE}
\begin{equation}
    \label{eq:rmse_detectors}
    \text{RMSE (detectors)}\coloneqq \left( \dfrac{1}{|\omega|}\sum_{i=0}^t \sum_{j=0}^l \left(\zeta_{i,j}^{\text{obs}, \omega} - \zeta_{i,j}^{\text{sim}, \omega} \right)^2 \right)^{1/2}, \zeta = q, v, \ \text{or} \ o.
\end{equation}
The idea of RMSE of macroscopic fields echos validation error, where the simulated trajectories are processed into macroscopic flow, speed and densities according to Edie's definition~\cite{edie1958traffic}, and are compared against the ones computed from the ground truth trajectories:

\begin{equation}
    \label{eq:rmse_macro}
    \text{RMSE (macroscopic fields)}\coloneqq \left( \dfrac{1}{|\Omega|}\sum_{i=0}^T \sum_{j=0}^L \left(\zeta_{i,j}^{\text{obs},\Omega} - \zeta_{i,j}^{\text{sim},\Omega} \right)^2 \right)^{1/2}, \zeta = q, v, \ \text{or} \ \rho.
\end{equation}
where $\Omega = [0,L] \times [0,T]$ defines the entire spatiotemporal domain of the simulation. The temporal resolution is 10s, and spatial resolution is 10m. $\rho=q/v$ is traffic density. These macroscopic fields serve as a higher-fidelity measurements than the stationary detectors, and provide a complete picture of the traffic characteristics from the entire spatial-temporal domain of the simulation.

\subsubsection{Results and Discussion}

The result of the synthetic experiment is summarized in Table~\ref{tab:synth_rmse}. Overall all of the 9 experiments can reproduce qualitatively the macroscopic features much better compared to suing the default parameters.

\begin{table}[!ht]
\resizebox{\textwidth}{!}{%
\begin{tabular}{@{}lccccccccccc@{}} 
\toprule
& & & \multicolumn{9}{c}{Calibrated parameters in each Exp} \\ \cmidrule(l){4-12}
Model features                 & Parameter & Calibration range & 1.a & 1.b & 1.c & 2.a & 2.b & 2.c & 3.a & 3.b & 3.c \\ \midrule
\multirow{5}{*}{Car-following} & $v_f$          &  [30.0,35.0]    & 30.48 & 30.50 & 32.62 & 32.0* &  32.0* &  32.0*  & 31.45 & 31.61 & 30.53\\
                               & $s_j$          &  [1.0,3.0] & 2.72 & 2.86 & 2.45 & 2.5* &2.5* & 2.5*  & 1.87 & 2.46 & 2.80\\
                               & $\tau $        & [0.5,2.0]   & 1.45 & 1.39 & 1.39 & 1.0* & 1.0* & 1.0*  & 1.40 & 1.45 & 1.37\\
                               & $a$            &  [1.0,4.0] & 1.10 & 1.11 & 1.00 & 2.6* & 2.6* & 2.6*  &  2.24 & 1.62 & 2.45\\
                               & $b$            & [1.0,3.0]  & 2.15 & 1.98 & 2.91 & 4.5* & 4.5* & 4.5* & 2.51 & 2.49 & 2.43\\ \midrule
\multirow{6}{*}{Lane-change}   & lcStrategic    & [0.0,5.0]  & 1.0* & 1.0* & 1.0* & 1.05 & 0.48 & 0.39  &  0.86 &  1.41 & 1.34 \\
                               & lcCooperative  & [0.0,1.0] & 1.0* & 1.0* & 1.0* & 0.86 & 0.86 & 0.74  & 0.98&  1.00& 1.00 \\
                               &  lcAssertive   & [0.0,5.0] & 1.0* & 1.0* & 1.0* & 0.39 & 0.19 & 0.18 & 0.43 & 0.55 &0.35 \\
                               &  lcSpeedGain   & [0.0,5.0] & 1.0* & 1.0* & 1.0* & 0.77 & 4.29 & 2.90  &1.14 & 3.76 & 1.90 \\
                               &  lcKeepRight   & [0.0,5.0] & 1.0* & 1.0* & 1.0* & 4.40 & 1.65 & 2.60  & 4.03 & 0.36 & 0.73 \\ 
                               \bottomrule
\end{tabular}}
\caption{Calibrated parameters for each experiment. * indicates that the values are fixed at default and are not subject to calibration.}
\label{tab:synth_calibrated_parameters}
\end{table}

\begin{table}[]
\centering
\begin{tabular}{@{}lllllll@{}}
\toprule
             & \multicolumn{3}{c}{RMSE at detector locations}                    & \multicolumn{3}{c}{RMSE of macroscopic fields}                  \\ \cmidrule(l){2-4} \cmidrule(l){5-7}
Exp. No.     &  $\q$ [vph] &  $\mathbf{v}$ [mph] &  $\mathbf{\rho}$ [vpm] &  $\q$ [vph] &  $\mathbf{v}$ [mph] &  $\mathbf{\rho}$ [vpm] \\ \midrule
Ground truth & 0         & 0                 & 0                &      0     &            0       &         0             \\
Default      & 476.06      &   11.40        &     14.03         &    1194.36     &     27.56         &      162.34        \\
1.a          &   154.2   &      5.40         &    24.91         &  493.79  &     7.03        &       82.91         \\
1.b          &  150.71       &   5.74         &     17.89            &   476.48   &     7.02        &        82.33        \\
1.c          &   148.01     &   7.60          &   28.79             &    553.71   &      13.62       &      101.91       \\
2.a          &   381.09     &    15.96       &    12.91          &  568.94     &       9.20       &      78.40             \\
2.b          &   387.36 &     12.12         &   12.30               &  609.71     &      17.71        &    87.40            \\
2.c          &     356.27  &     9.39          &      19.79            &  609.84    &     11.00       &    68.97         \\
3.a          &    114.01    &  9.61            &     16.16             &  410.42     &    5.37        &   69.72            \\
3.b          & \textbf{110.16}     &   \textbf{5.09}    &  \textbf{6.79}       &  \textbf{398.51}    &   \textbf{4.23}   &  \textbf{61.40}             \\
3.c          &  111.28    &  7.20           &   7.73            &  419.77     &   8.93          &   64.02           \\ \bottomrule
\end{tabular}
\caption{Synthetic experiment results. vph: vehicles per hour; mph: miles per hour; vpm: number of vehicles per mile.}
\label{tab:synth_rmse}
\end{table}

\begin{figure}[!htb]
  \centering
  \includegraphics[width=\textwidth]{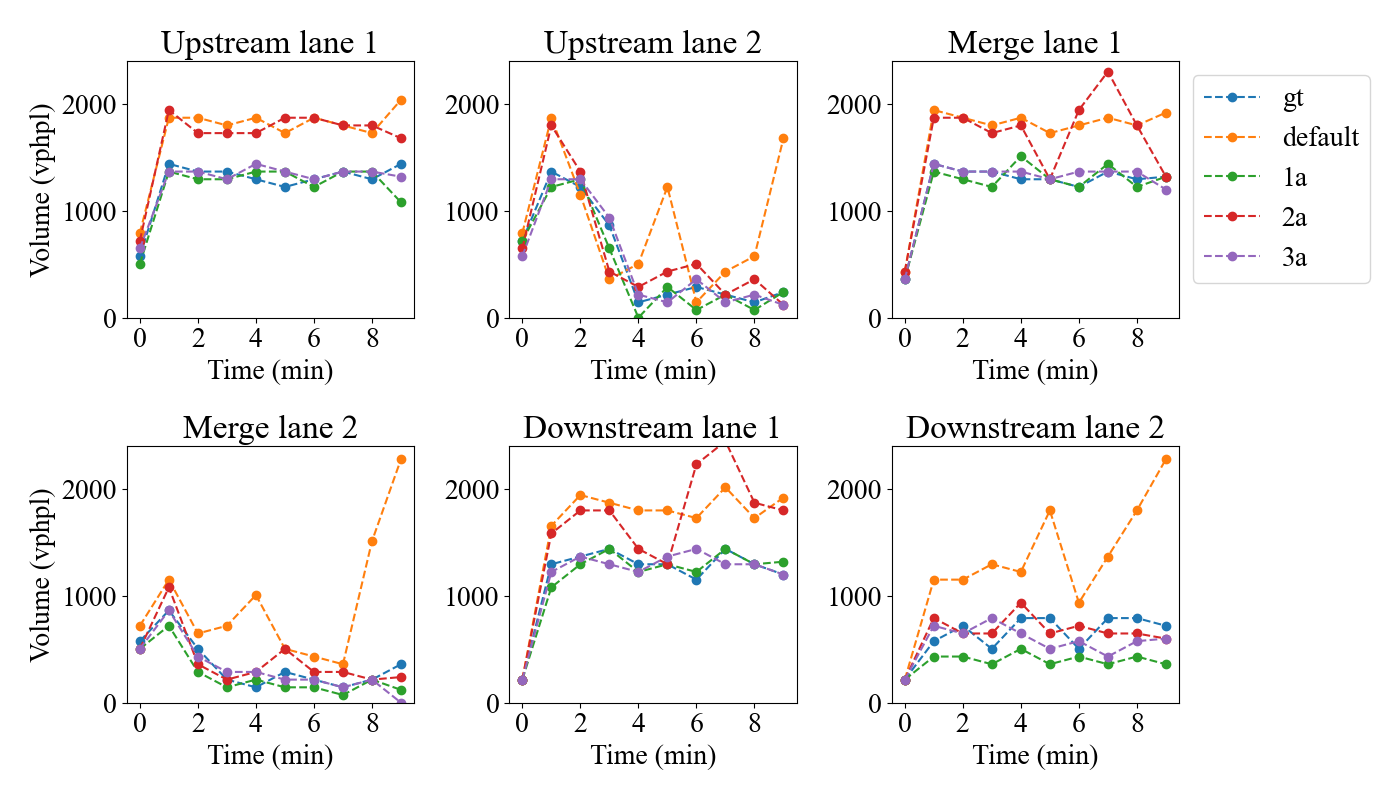}
   \caption{A lane-by-lane comparison of flow (vphpl: vehicles per hour per lane), for Exp.1.a, 2.a and 3.a.}
  \label{fig:synth_detector_flow}
\end{figure}

\begin{figure}[!htb]
  \centering
  \includegraphics[width=\textwidth]{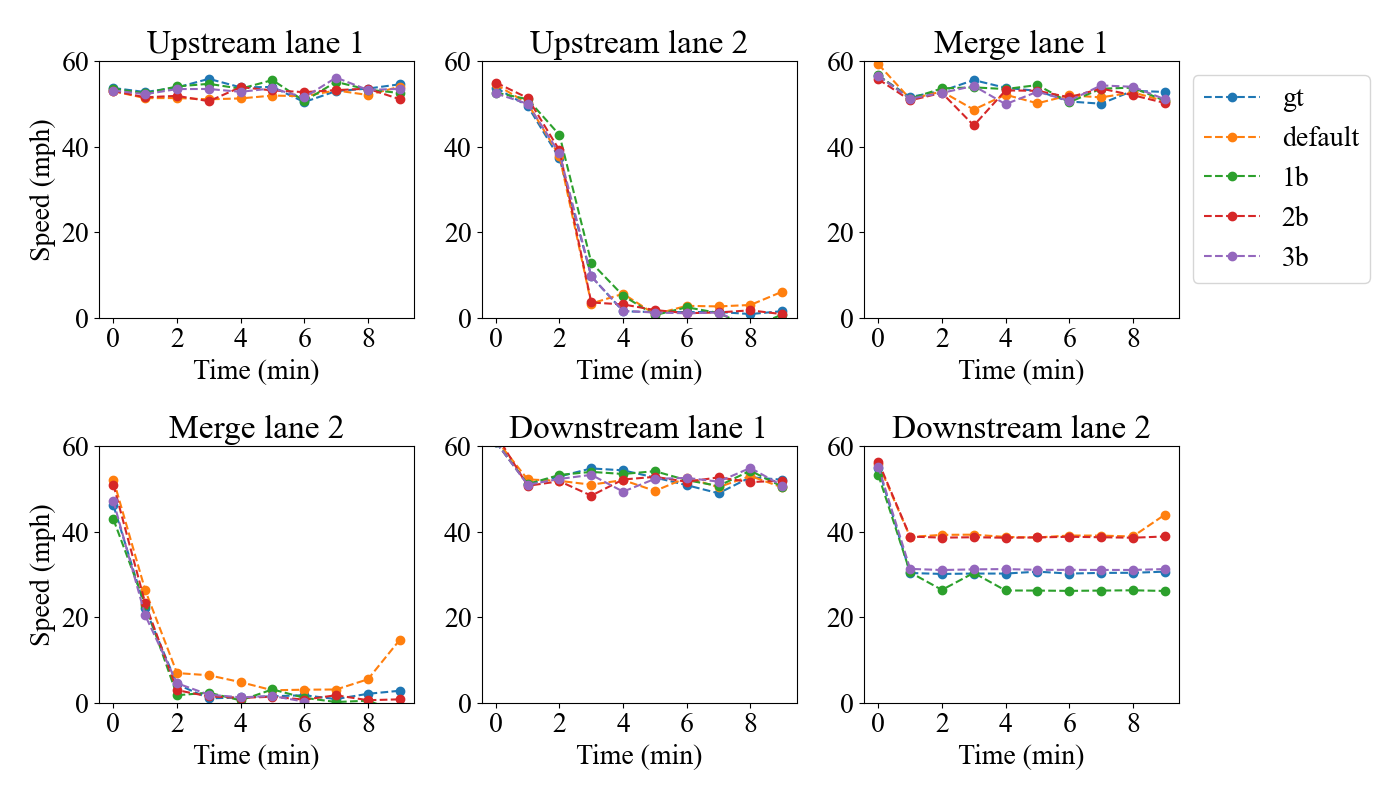}
  \caption{A lane-by-lane comparison of speed (mph), for Exp.1.b, 2.b and 3.b.}
  \label{fig:synth_detector_speed}
\end{figure}

\begin{figure}[!htb]
  \centering
  \includegraphics[width=\textwidth]{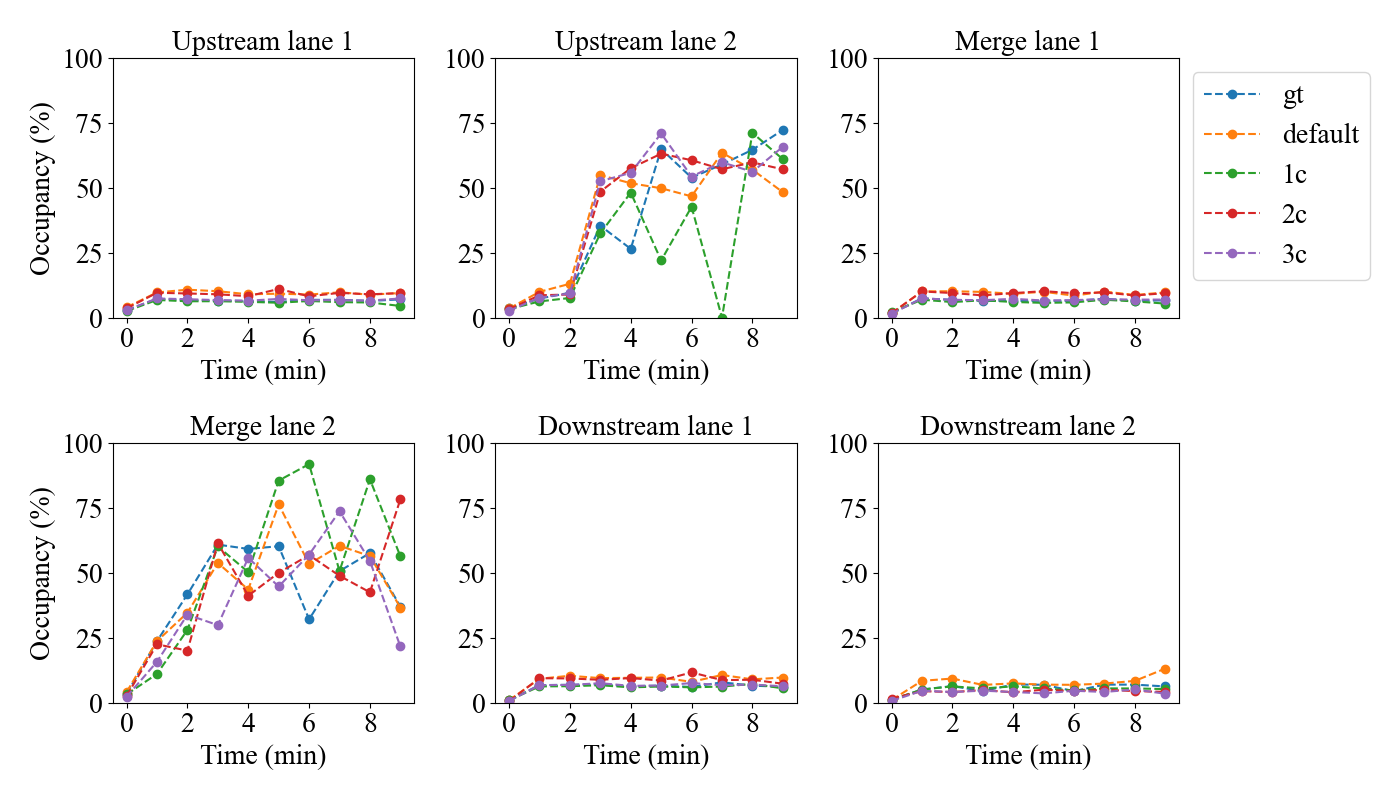}
   \caption{A lane-by-lane comparison of occupancy (in \%), for Exp.1.c, 2.c and 3.c.}
  \label{fig:synth_detector_occ}
\end{figure}

The results provide a few key insights. First, the effects of car-following and lane-change parameters are highly coupled. This is evident in three key observations. Firstly, from the calibrated values in Table~\ref{tab:synth_calibrated_parameters}, the optimal car-following and lane-change parameters identified in Exp.3.a, 3.b and 3.c differ from those obtained when calibrating each parameter independently in Exp 1's and 2's. Secondly, Figure~\ref{fig:synth_detector_flow}-\ref{fig:synth_detector_occ} illustrate that calibrating lane-change parameters alone (with $\thetacf$ set to default) does not always minimize lane-specific errors and can sometimes yield worse results, as seen in Figure~\ref{fig:synth_detector_flow}, for upstream lane 1 and downstream lane 1 in Exp. 2.a, where lane-change parameter calibration deviates most from the ground truth. The best results, in terms of lower RMSE at lane-specific detector locations, are achieved when $\thetacf$ is calibrated either alone (Exp 1.a) or in conjunction with $\thetalc$ (Exp 3.a). Finally, Table~\ref{tab:synth_rmse} indicates that joint calibration of $\thetacf$ and $\thetalc$ (Exp. 3) reduces RMSEs both at the detector locations and across macroscopic fields, as compared to independent calibration (Exp. 1 and 2).


Next we examine the choice of measurements on the calibration performance in the single-objective optimization formulation~\eqref{eq:opt_rmse}. From Table~\ref{tab:synth_rmse}, we observe that when calibrating $\thetacf$ alone, the lowest RMSE on macroscopic fields is achieved with the speed measurement (Exp. 1.b). Similarly, when jointly calibrating $\thetacf$ and $\thetalc$, the best result also corresponds to using the speed measurements (Exp. 3.b). However, this pattern does not hold when calibrating $\thetalc$ alone (Exp.2.a-2.c). This discrepancy is likely because the lane-change parameters have minimal direct impact on the simulated speed.

\begin{figure*}[!htb]
\begin{minipage}{\textwidth} 
    \includegraphics[width=\linewidth]{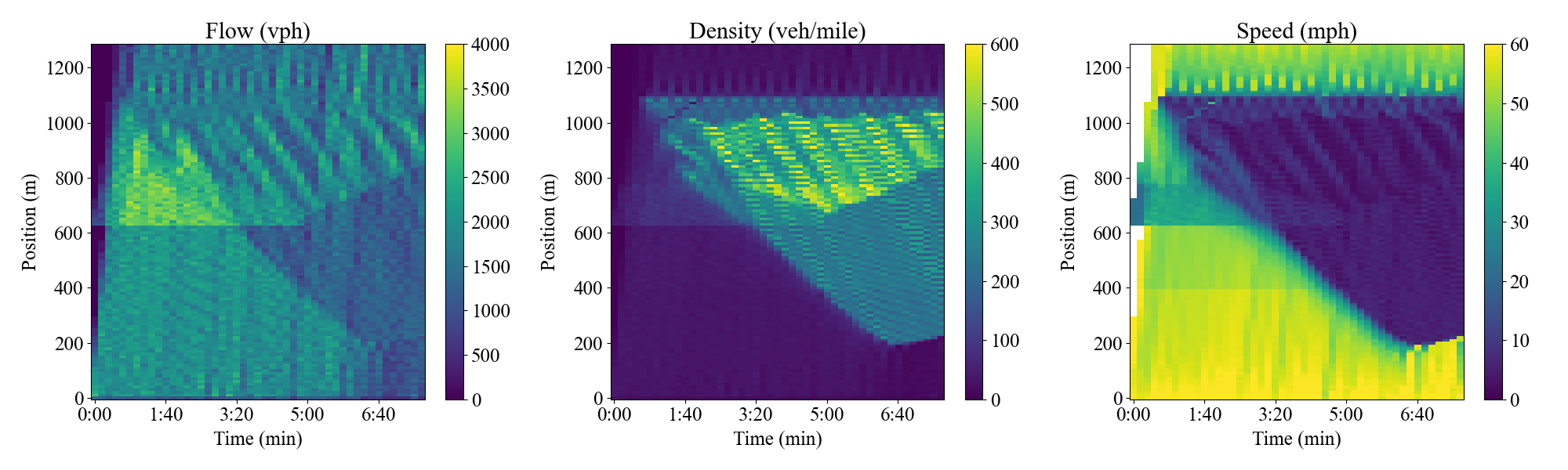}
    \subcaption{Macroscopic quantities with ground truth $\theta_{\text{CF}}$ and $\theta_{\text{LC}}$ parameters shown in Table~\ref{tab:parameters}. }
    \label{fig:macro_gt}
\end{minipage}    
\hspace{\fill}  
\begin{minipage}{\textwidth} 
    \includegraphics[width=\linewidth]{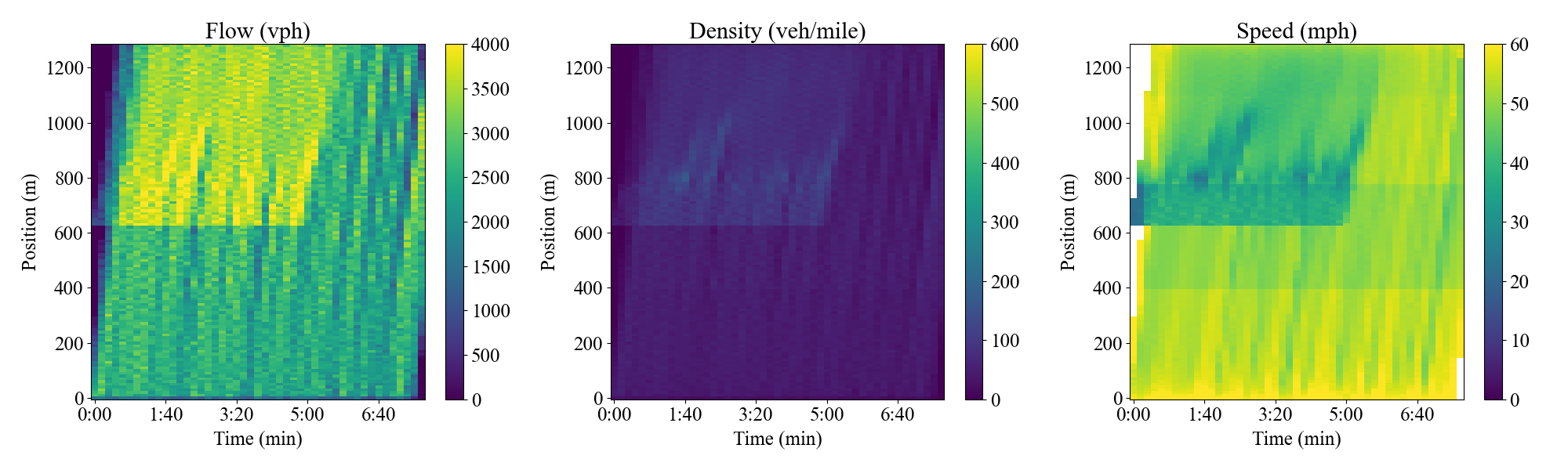}
    \subcaption{Macroscopic quantities using default $\theta_{\text{CF}}$ and $\theta_{\text{LC}}$ parameters shown in Table~\ref{tab:parameters}.}
     \label{fig:macro_default}
\end{minipage}  
\hspace{\fill}  
\begin{minipage}{\textwidth} 
    \includegraphics[width=\linewidth]{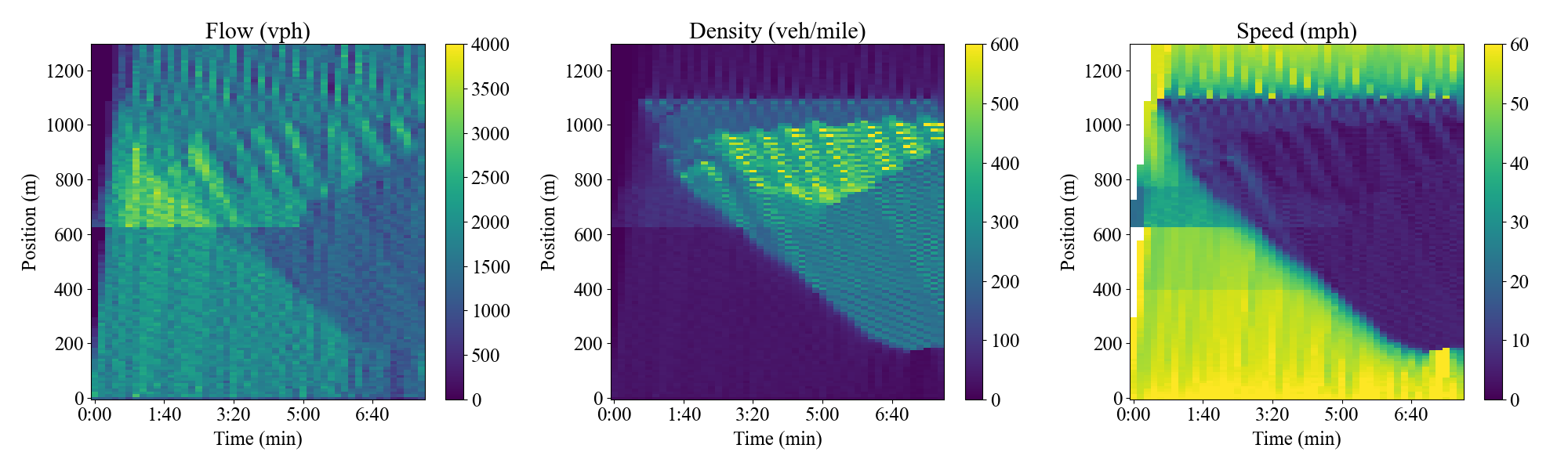}
    \subcaption{Macroscopic quantities using the best $\theta_{\text{CF}}$ and $\theta_{\text{LC}}$ parameters shown in Table~\ref{tab:parameters}.}
     \label{fig:macro_3b}
\end{minipage} 
\caption{Macroscopic quantities using the ground-truth and default (uncalibrated) parameters.}
\label{fig:macro_gt_default}
\end{figure*}

\begin{figure}[!htb]
  \centering
  \includegraphics[width=\textwidth]{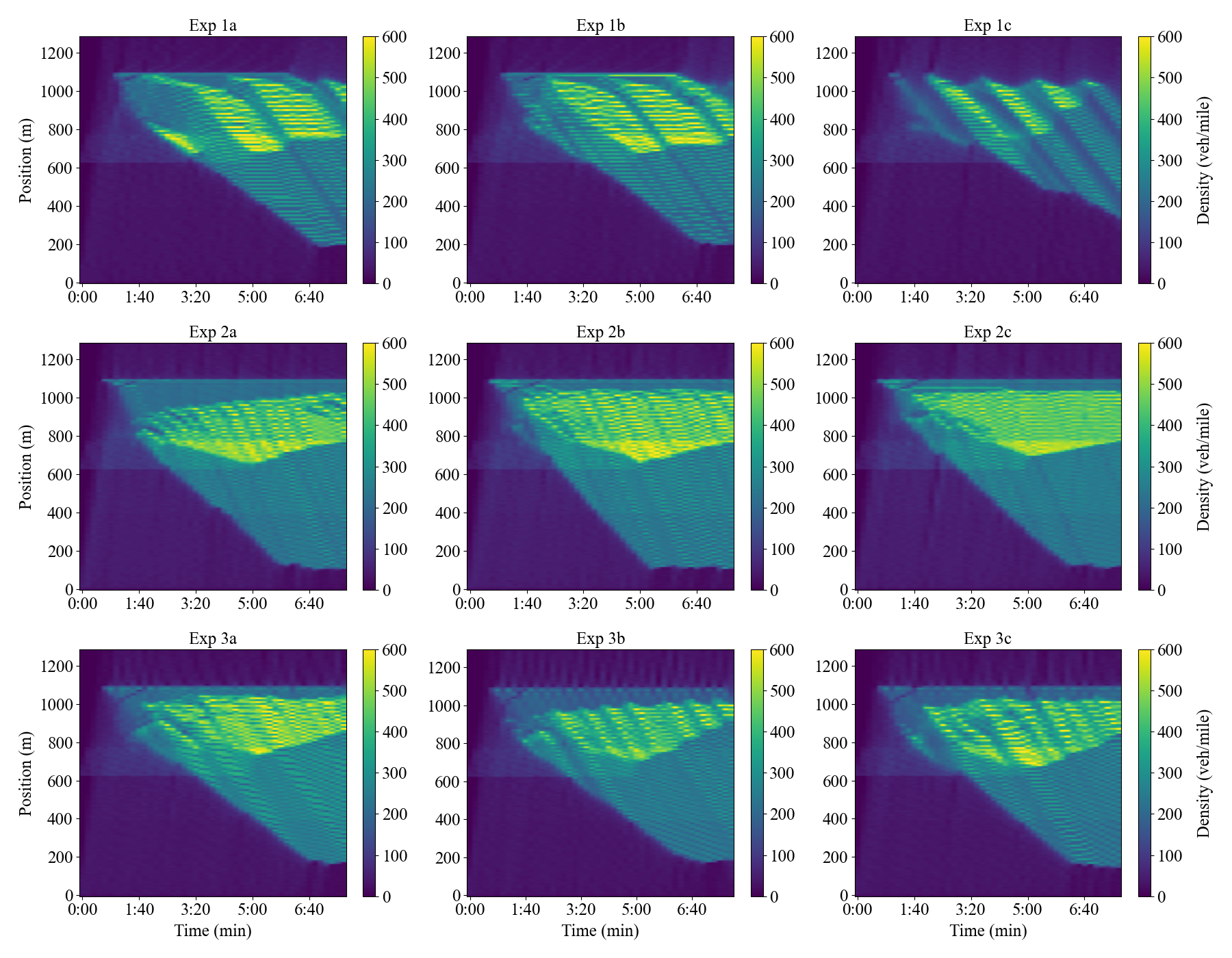}
   \caption{Spatiotemporal density map (vpm) for all 9 experiments.}
  \label{fig:synth_macro_density}
\end{figure}

\begin{figure}[!htb]
  \centering
  \includegraphics[width=\textwidth]{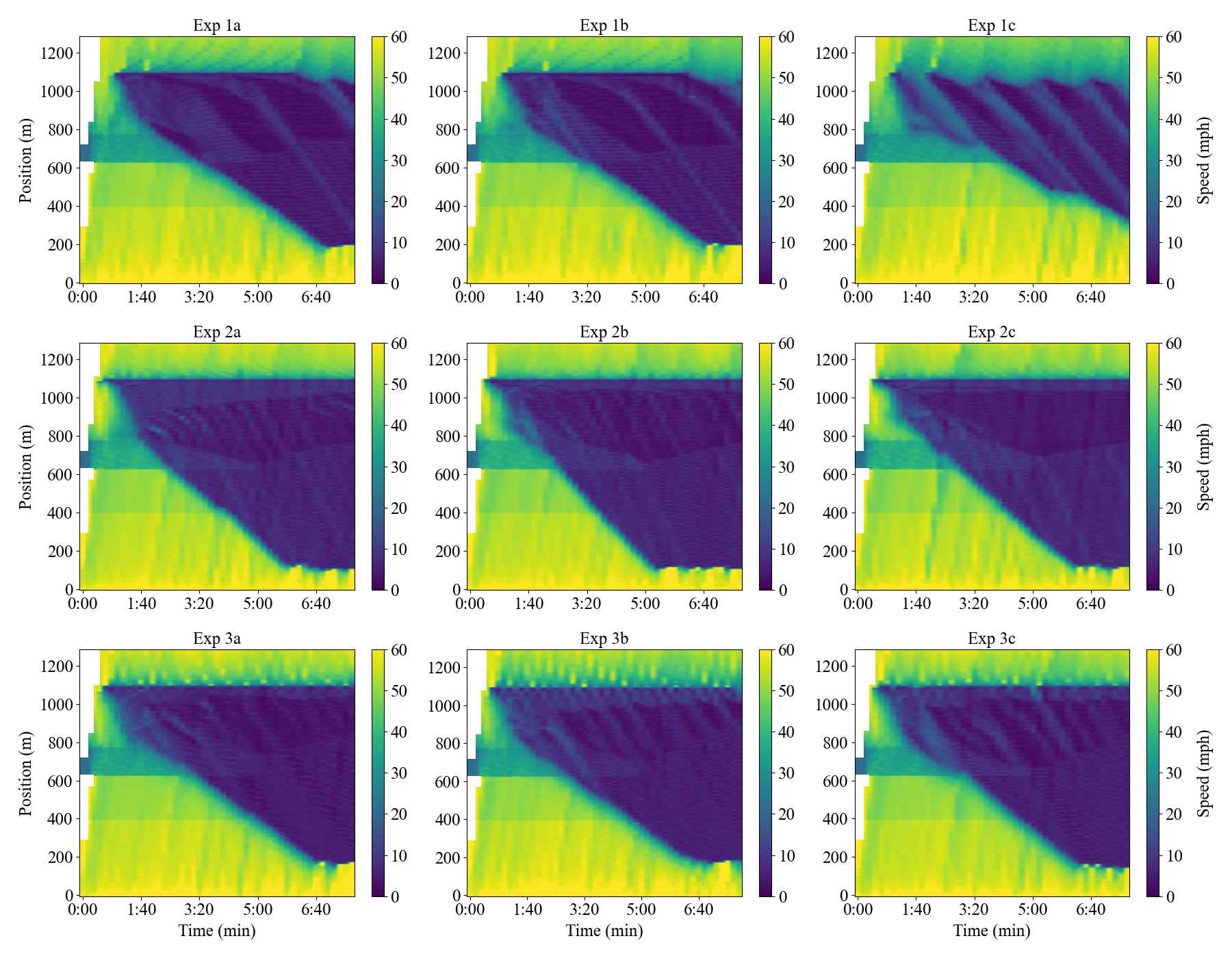}
   \caption{Spatiotemporal speed map (mph) for all 9 experiments.}
  \label{fig:synth_macro_speed}
\end{figure}

All calibration configurations can qualitatively reproduce macroscopic traffic patterns, while the default (uncalibrated) parameters fail to capture the congestion caused by on-ramp merging, evident when comparing the ground truth macroscopic fields (Figure~\ref{fig:macro_gt}) with the default parameters' simulation (Figure~\ref{fig:macro_default}). This issue arises because the default acceleration and deceleration parameters are significantly higher than the ground truth values, causing queues to dissipate too quickly downstream. Figures~\ref{fig:synth_detector_flow} to \ref{fig:synth_macro_speed} compare macroscopic patterns from each experiment. Generally, all experiments capture the congestion upstream of the merging zone (900m). Notably, Exp 3 best represents the shape of the jam front (see density Figure~\ref{fig:synth_macro_density} and speed Figure~\ref{fig:synth_macro_speed}), with Exp 3.b and 3.c particularly well capturing the stop-and-go waves between 750m and 1000m, as seen in the ground truth. These results emphasize the need for microsimulation model calibration, as collective driving behaviors significantly impact overall congestion patterns.

These synthetic experiments also provide insights into configuring the DE optimization algorithm. The results were generated with a population size between 10 and 20 and a maximum of 100 iterations. The program terminated after approximately 15,340 objective function evaluations per experiment. However, none of the runs successfully terminated, as the convergence criteria were not met before reaching the maximum iterations.

\subsection{Real-World Experiment: I-24 Corridor}

\begin{figure}
  \centering
  \includegraphics[width=\textwidth]{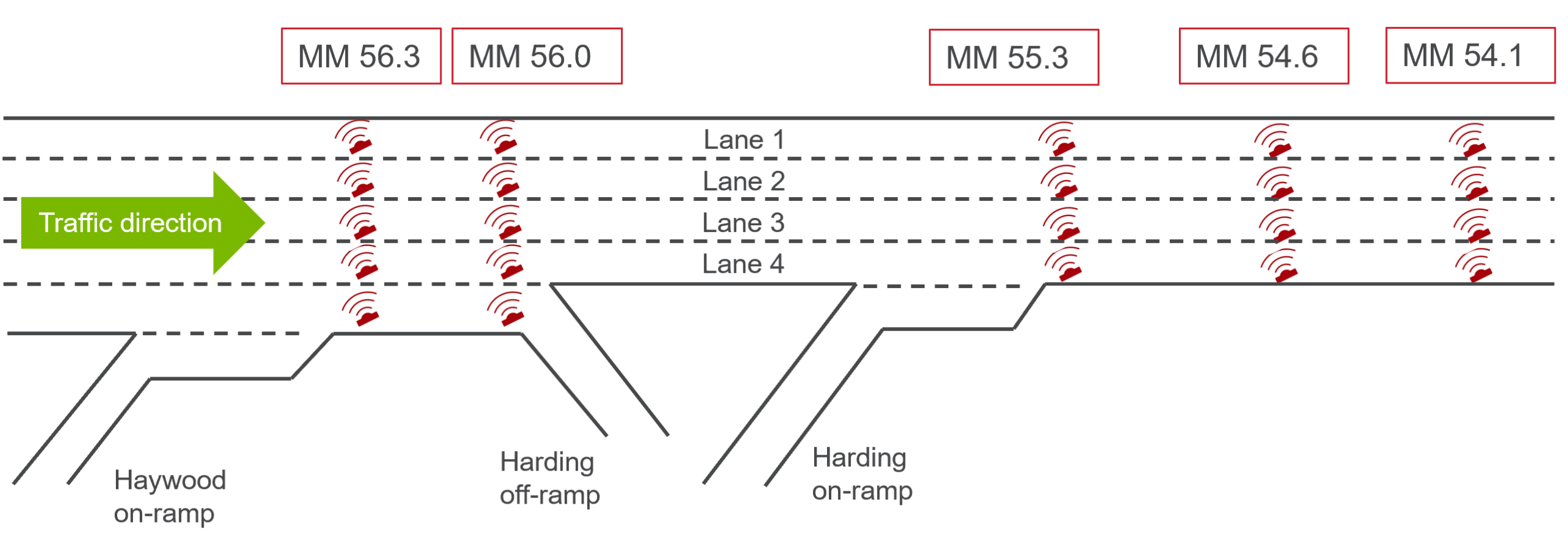}
  \caption{The I-24 westbound segment (3.5 miles) set up in SUMO as a freeway segment with two on-ramps and one off-ramp. Traffic flows in the direction of decreasing mile markers. Red icons indicate the  locations where lane-specific traffic information is available from RDS sensors.}
  \label{fig:i24_corridor}
\end{figure}

The second microsimulation calibration scenario in this paper focuses on a 3.5-mile segment of the Interstate-24 corridor, between mile markers (MM) 54.1 and 57.6, westbound. This segment was selected because it is where recurring congestion typically begins in the morning peak, due to high on-ramp demand and its complex geometry (see Figure~\ref{fig:i24_corridor}). Downstream of MM 54.5 is generally free-flowing, therefore characterizing an active bottleneck~\cite{cassidy1999some}.  The time range for this study is from 5:00 to 8:00 AM on a typical workday morning (November 13, 2023), when congestion accumulates during rush hour.

\subsubsection{RDS Data and Preprocessing}
The Tennessee Department of Transportation maintains a surveillance system to collect traffic data across various highway facilities in the state~\cite{tdot}. The data, aggregated in 30-second intervals, provides lane-by-lane occupancy, counts, and speed information. This data was downloaded from the I-24 MOTION project website~\cite{gloudemans202324}.

Pre-processing is necessary to address inconsistencies due to miscounts, communication failures, and missing sensors, especially at on- and off-ramps. For instance, the Harding on-ramp (MM 55.6) was missing, so demand was estimated based on downstream counts using flow conservation, and the temporal pattern was inferred from the Haywood on-ramp (MM 56.7). The data was used to set the upstream demand (MM 57), the two on-ramps, and the turning rate at off-ramp MM 56.0. Occupancy and counts aggregated in 5-minute intervals from detectors at MM 54.1, 54.6, 55.3, 56.0, and 56.3 were used for calibration. Aggregating the data at 5-minute intervals helps eliminate high-frequency noise while preserving sufficient time-varying information.

\begin{figure}[!ht]
  \centering
  \includegraphics[width=\textwidth]{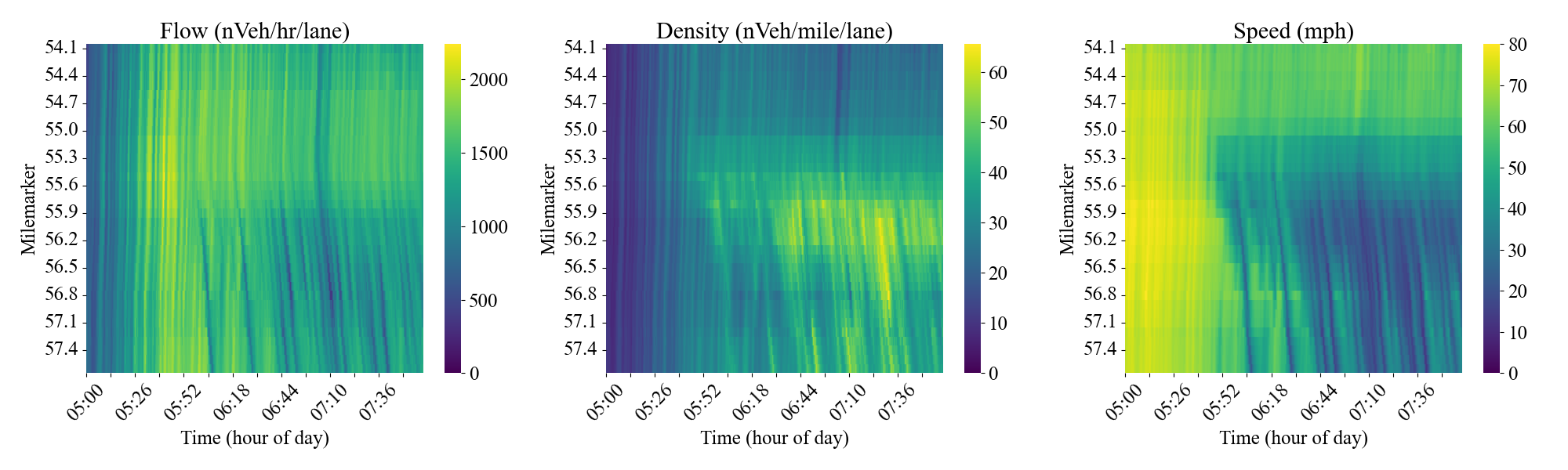}
   \caption{The reconstructed macroscopic quantities from stationary RDS data using ASM.}
  \label{fig:ASM}
\end{figure}

In order to get complete spatiotemporal traffic data from those 5 stationary sensors, the Adaptive Smoothing Method (ASM)~\cite{treiber2011reconstructing} was applied. It is a technique to reconstruct smooth, continuous state estimates of flow, speed, and density by adjusting its smoothing parameters dynamically based on the local data density and traffic conditions. For a recent application of ASM, please refer to~\cite{ji2024virtual}. The ASM-processed traffic states can be visualized in Figure~\ref{fig:ASM}, which is used as a validation dataset to evaluate the calibration performance of the entire spatiotemporal region. The temporal resolution is 30s, and spatial resolution is 0.1-mile. Same discretization is applied to the simulated trajectory data for comparison.

\subsubsection{Simulation Setup}
The SUMO network file for this segment is obtained from OpenStreetMap~\cite{OpenStreetMap} with a few manual corrections on the lane configurations. Similar to the synthetic experiments, we consider 9 combinations of parameters and measurements for calibration (Table~\ref{tab:two_experiments}). The default car-following and lane-change parameters are set the same as the ones specified in Table~\ref{tab:parameters}. We follow the same evaluation procedures as in the previous experiments, where the RMSE at detector locations and the RMSE for the entire spatiotemporal macroscopic fields are computed. The benchmark dataset to evaluate RMSE of macroscopic quantities is the reconstructed data from RDS using ASM. Demands for each potential route from each origin (upstream and on-ramps) to each destination (off-ramp if origin is upstream and the downstream section). The total demand at each origin was divided into each destination based on the turning ratio at the off-ramp. Since the input demand already follows a Poisson distribution, the temporal distribution of the demand was binned into steps of 30 minutes too avoid extreme variability in demand.

\subsubsection{Results and Discussion}
\label{sec:discussion}
\begin{table}
\resizebox{\textwidth}{!}{%
\begin{tabular}{@{}lccccccccccc@{}} 
\toprule
& & & \multicolumn{9}{c}{Calibrated values from Exp} \\ \cmidrule(l){4-12}
Model features                 & Parameter & Bounds & 1.a & 1.b & 1.c & 2.a & 2.b & 2.c & 3.a & 3.b & 3.c \\ \midrule
\multirow{5}{*}{Car-following} & $v_f$          &  [25.0,40.0] & 36.87 & 31.53 & 30.93 & 32.0* &  32.0* &  32.0*  & 28.04 & 40.23 & 27.34\\
                               & $s_j$          &  [0.5,3.0] & 2.26 & 1.86 & 0.58 & 2.5* &2.5* & 2.5*  & 1.67 & 2.32 & 0.74\\
                               & $a$            &  [1.0,4.0] &  3.91 & 1.07 & 1.59 & 2.6* & 2.6* & 2.6*  & 3.98 & 1.78 & 1.29\\
                               & $b$            & [1.0,4.0]   &  1.68 & 3.89 & 1.00 & 4.5* & 4.5* & 4.5* & 1.00 & 3.89 & 1.00\\ \midrule
                               & $\tau $        &  [0.5,2.0]  & 0.82 & 1.79 & 2.56 & 1.0* & 1.0* & 1.0*  & 1.19 & 1.85 & 2.03\\
\multirow{6}{*}{Lane-change}   & lcStrategic    & [0.0,5.0]  & 1.0* & 1.0* & 1.0* & 0.84 & 0.09 & 0.004  & 4.01 &  0.09 & 1.59 \\
                               & lcCooperative  & [0.0,1.0] & 1.0* & 1.0* & 1.0* &0.90 & 0.39 & 0.06  & 0.55 &  0.34& 0.92 \\
                               &  lcAssertive   & [0.0,5.0] & 1.0* & 1.0* & 1.0* & 4.84 & 2.47 & 2.81 & 4.56 & 2.71 &4.52 \\
                               &  lcSpeedGain   & [0.0,5.0] & 1.0* & 1.0* & 1.0* & 1.91 & 4.93 & 2.90  &4.59 &  1.04 & 4.74 \\
                               &  lcKeepRight   & [0.0,5.0] & 1.0* & 1.0* & 1.0* & 0.09 & 0.17 & 0.56  & 0.05 & 0.24 &0.13 \\ 
                               \bottomrule
\end{tabular}}
\caption{Calibrated parameters for each experiment. * indicates that the values are fixed at default and are not subject to calibration.}
\label{tab:i24_calibrated_parameters}
\end{table}

\begin{table}[]
\centering
\begin{tabular}{@{}lllllll@{}}
\toprule
             & \multicolumn{3}{c}{RMSE at detector locations}              & \multicolumn{3}{c}{RMSE of macroscopic fields}                \\ \cmidrule(l){2-4} \cmidrule(l){5-7}
Exp. No.     & $\q$ [vphpl] & $\mathbf{v}$ [mph] & $\mathbf{\rho}$ [veh/mile/lane] & $\q$ [vphpl] & $\mathbf{v}$ [mph] & $\mathbf{\rho}$ [veh/mile/lane] \\ \midrule
1.a          & 229.27       & 29.37             & 18.38                            & 309.07       & 29.92             & 33.02                            \\
1.b          & 491.80       & 14.78             & 18.61                            & 519.97       & 19.92             & 32.95                            \\
1.c          & 415.15       & 17.69             & 16.74                            & 435.31       & 18.68             & 32.97                            \\
2.a          & 364.19       & 24.46             & 19.61                            & 491.90       & 23.60             & 33.02                            \\
2.b          & 528.43       & 11.61             & 15.76                            & 560.39       & 16.65             & 32.97                            \\
2.c          & 430.95       & 11.90             & 14.06                            & 514.25       & 17.45             & 32.99                            \\
3.a          & \textbf{212.62} & 18.80             & 15.56                            & \textbf{270.85} & 18.50             & 33.01                            \\
3.b          & 304.98       & \textbf{8.51}      & \textbf{11.30}                   & 379.41       & \textbf{15.34}    & 32.98                            \\
3.c          & 290.21       & 22.73             & 27.57                            & 316.75       & 22.52             & \textbf{32.92}                   \\ \bottomrule
\end{tabular}

\caption{Calibration performance for I-24 corridor. vphpl: vehicles per hour per lane; vpmpl: vehicles per mile per lane.}
\label{tab:i24_rmse}
\end{table}

\begin{figure}[!ht]
  \centering
  \includegraphics[width=\textwidth]{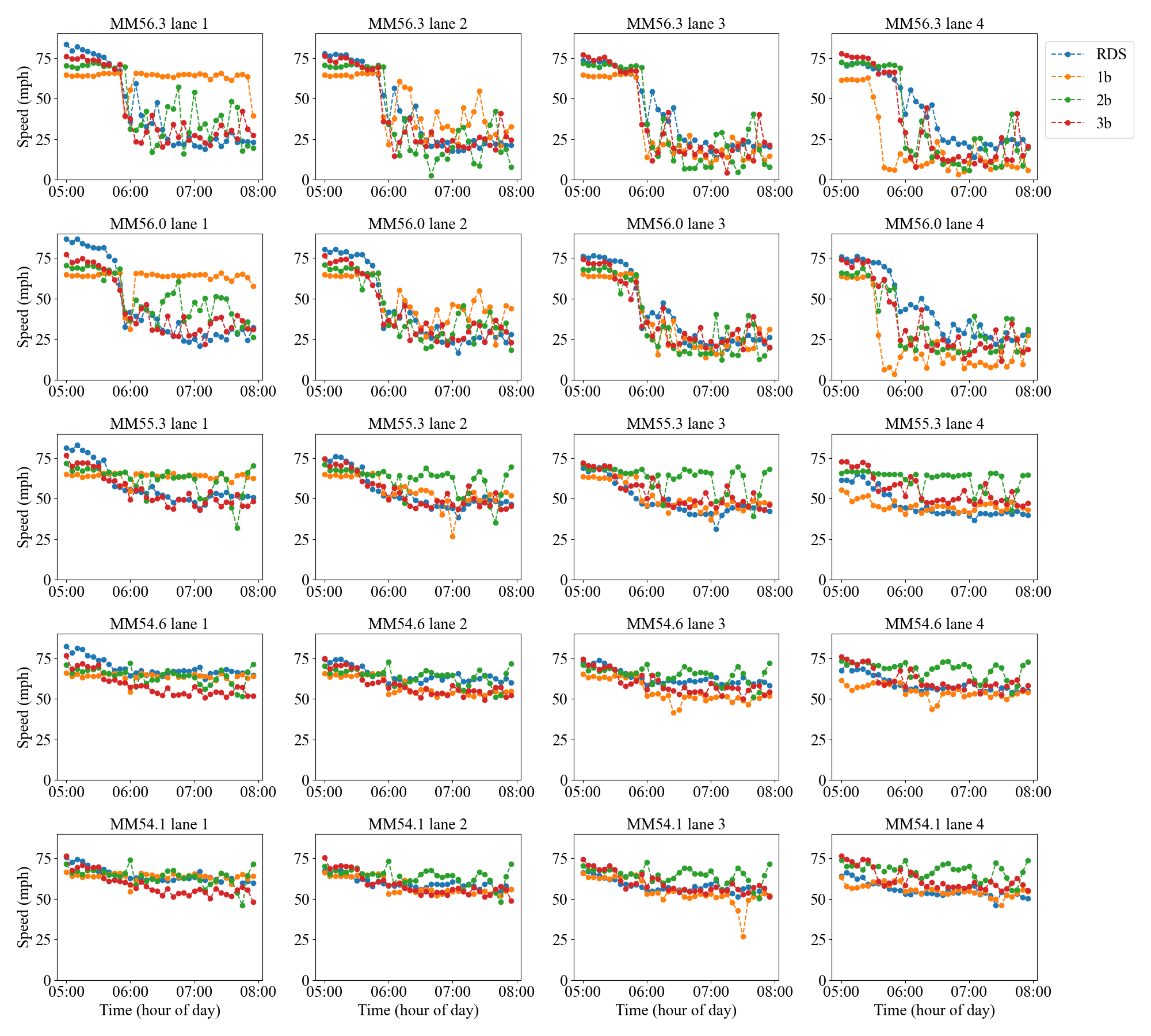}
   \caption{Lane-by-lane speed comparison (in miles per hour) for Exp.1.b., 2.b and 3.b.}
  \label{fig:i24_detector_speed}
\end{figure}

We observe a similar trend as in the synthetic scenario: having more degrees of freedom for parameter calibration produces better results. Table~\ref{tab:i24_rmse} provides quantitative evidence: Exp 3 yields lower RMSE of macroscopic fields compared to Exp 1 and 2. A lane-specific comparison with RDS data (Figure~\ref{fig:i24_detector_speed}) suggests that Exp 3.b (jointly calibrating $\thetacf$ and $\thetalc$) better captures the onset of the slowdown. In contrast, calibrating $\thetalc$ alone (Exp 2.b) overestimates the speed because the default free-flow speed $v_f$ is set too high. Calibrating only $\thetacf$ (Exp 1.b) fails to capture the tactical and strategic lane-changing behavior near MM 56.3, as cars tend to accumulate near the right-lane merging zone and are less likely to maneuver left.

\begin{figure}
  \centering
  \includegraphics[width=\textwidth]{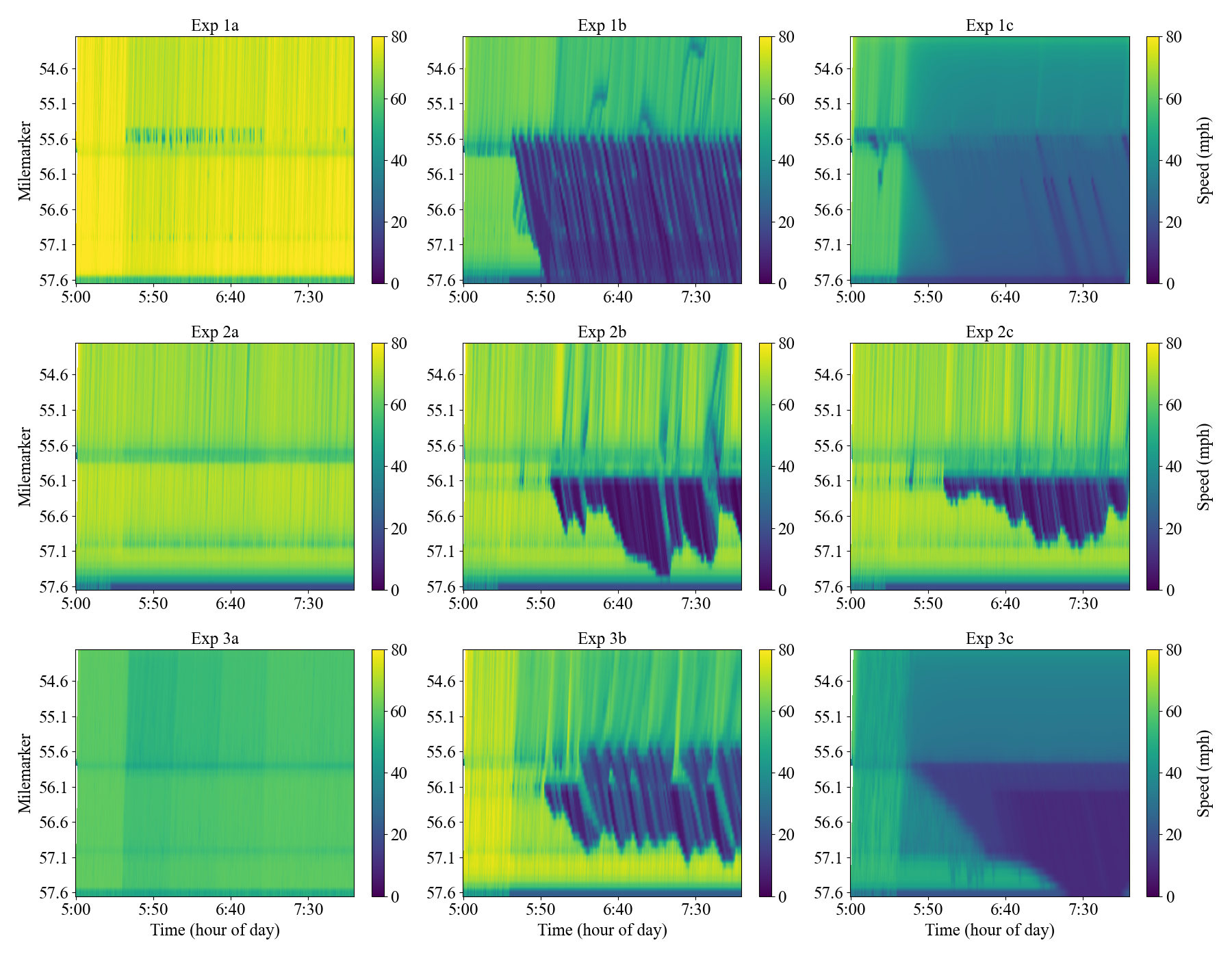}
   \caption{Spatiotemporal speed map (mph) for all 9 experiments in the I-24 scenario.}
  \label{fig:macro_i24_speed}
\end{figure}

While the synthetic scenario suggests that speed measurements provide the best calibration results, this is not as evident in the I-24 scenario. From Table~\ref{tab:i24_rmse}, Exp 2.b and 3.b show that speed measurements help recover macroscopic fields better than flow and density measurements. However, this is not true for Exp 1.b, where density measurements seem more effective. Despite this, the spatiotemporal speed map produced from the calibrated results in all nine experiments (Figure~\ref{fig:macro_i24_speed}) indicates that using speed or density measurements can generate some level of congestion from the merging zones. This pattern is not captured at all when using flow measurements.

Figure~\ref{fig:macro_i24_speed} also shows that Exp 1 (calibrating $\thetacf$ only) and Exp 3 (calibrating both $\thetacf$ and $\thetalc$) correctly identify the onset of congestion. The RDS data suggests that congestion starts at around MM 55.3, just downstream of the Harding on-ramp, which is accurately captured in Exp 1 and 3. However, Exp 2 incorrectly suggests that congestion starts at the upstream Haywood on-ramp.

This scenario suggests that multiple criteria should be considered when evaluating the results. Figure~\ref{fig:i24_detector_speed} offers direct lane-by-lane comparison, while Table~\ref{tab:i24_rmse} offers a quantitative evaluation of the calibrated results' performance, especially the models' predictive capability across the entire spatiotemporal range. Figure~\ref{fig:macro_i24_speed} provides a clear visualization of congestion and traffic wave patterns over the domain.

We see that Exp 1.b, although not scoring the lowest RMSE, performs the best at reproducing the location of the congestion onset, the queue propagation speed, and even the frequency of the traffic waves. In contrast, Exp 2 suggests that congestion does not backpropagate long enough, and Exp 3 suggests that queues do not accumulate fast enough, even though Exp 3 best captures the lane-specific speed variation. It is challenging to determine the optimal parameter combination and measurements for the best calibration practice. Ultimately, the microsimulation serves as a testing platform for new connected and automated vehicles (CAV) technologies and traffic management strategies. For example, CAV control may prioritize traffic wave dissipation and fuel consumption improvement~\cite{delle2019feedback,stern2018dissipation,kreidieh2018dissipating,avedisov2020impacts}, while infrastructure control, such as variable speed limits, may focus more on lane-specific speed variation and overall travel delay due to congestion~\cite{chen2014variable,chen2015variable}. Application-specific criteria can be leveraged to guide the benchmark for microsimulation calibration.

The I-24 scenario presents more challenges compared to the synthetic scenario. The first challenge is the longer segment and extended time duration, which require extended simulation time. This increases the chances of errors accumulating over time and space, resulting in a more complex terrain for the objective function. Similar to the synthetic scenario, no optimization runs successfully terminate while satisfying predefined convergence criteria.

Additionally, there are many more driving behavior-related parameters in SUMO that could potentially alter the traffic patterns. Even with the selected car-following and lane-change parameters, their impact on the simulated traffic pattern is highly coupled, suggesting that joint parameter calibration is necessary. If more decision variables are considered, a higher-dimensional search space needs to be explored, making the problem highly complex. An iterative approach might be necessary, but it is still essential to capture the coupling effect of parameters.

Another challenge is that realistic driving behavior on I-24 (or any real-world situation) is highly heterogeneous, while in this work, we constrain the simulation to a homogeneous fleet (where every vehicle has the same behavioral parameters). This heterogeneity is more evident during slow traffic and in merging zones, which are present in this scenario. Representing driving heterogeneity itself is challenging and can significantly affect the overall traffic patterns~\cite{punzo2020two}, as driver heterogeneous reactions to deceleration waves are found to cause various traffic oscillation patterns~\cite{laval2010mechanism}.

Nevertheless, despite considering only limited driving behavior parameters and finding only a sub-optimal solution from the optimization problem, the results are still promising. In some cases, the onset and propagation of congestion can be reproduced in the microsimulation. Additionally, the lane-specific traffic patterns, especially in the right lanes near merging and diverging zones, can be faithfully represented.

\section{Conclusion and Future Work}
\label{sec:conclusion}
In this work, we aim to reproduce observed traffic congestion patterns on freeways using the SUMO microsimulation tool. We approach this as an optimization problem to calibrate driving behavior-related parameters, including car-following and lane-changing parameters, based on macroscopic traffic measurements from stationary sensors. This method has proven effective in synthetic scenarios. However, while we have obtained promising results, we also encounter challenges when applying the method to a real-world freeway segment using field measurements. Numerous experiments have been conducted to evaluate the impact of different model parameters and measurement choices on calibration performance.

Speed measurements are generally the most crucial data for calibrating microsimulation models. Both synthetic and real-world scenarios demonstrate that speed data alone can qualitatively reproduce congestion. In some cases, such as Exp 1.b from the I-24 scenario, speed measurements can accurately estimate the spatiotemporal range of congestion. Exp 3.b best recovers the lane-specific traffic patterns. In contrast, flow measurements are the least effective for model calibration. This ineffectiveness likely stems from their ambiguity; low flow rates can indicate either fast, low-density traffic or slow, high-density traffic. The congested scenarios considered in this work typically include segments with low flow rates.

As mentioned in the discussion (Section~\ref{sec:discussion}), we are interested in exploring and calibrating the model against a variety of application-specific performance measures, such as travel time, capacity drop at merging bottlenecks, and queue lengths. Additionally, the high-fidelity I-24 MOTION data (from MM 58.7 to MM 62.9) upstream of the studied segment can be leveraged. The detailed trajectory data will allow us to calibrate and validate the model with microscopic metrics, such as counts of lane-change events and speed and space-gap distributions. This data also provides critical insights into studying the propagation and dissipation of traffic waves, which complements the congestion emergence observed in this study.

It will also be interesting to assess the performance of calibrated models with different input and demand information from other days. As seen in~\cite{gloudemans202324}, traffic patterns on I-24 vary daily, although the locations where recurring congestion begins and the time range of morning rush hours are generally similar across days. We aim to determine whether a model calibrated with one day's data can predict the traffic pattern on a different day or if route choices and origin–destination (O-D) flows must be calibrated alongside the driving behavior parameters~\cite{toledo2004calibration, chu2003calibration}.

\newpage

\section{Acknowledgements}
The authors would like to thank the Tennessee Department of Transportation for providing the RDS data and Junyi Ji (Vanderbilt University) for providing the validation dataset for the I-24 corridor experiment in this paper. We are grateful to Jihun Han and Tyler Ard (Argonne National Laboratory) for the technical discussions on the optimization problem, and Yaozhong Zhang (Argonne) for contributing the SUMO OpenStreetMap.

 The submitted manuscript has been created by UChicago Argonne, LLC, Operator of Argonne National Laboratory (“Argonne”). Argonne, a U.S. Department of Energy Office of Science laboratory, is operated under Contract No. DE-AC02-06CH11357. This report and the work described were sponsored by the U.S. Department of Energy (DOE) Vehicle Technologies Office (VTO) under the Energy Efficient Mobility Systems (EEMS) Program, with support from EERE managers Avi Mersky, Erin Boyd and Alexis Zubrow.

\bibliographystyle{abbrv}
\bibliography{trb_template}
\end{document}